\documentclass[
10pt, twocolumn, prd, superscriptaddress
]{revtex4-1}
\usepackage{graphicx, hyperref, color, bm, amsmath, amssymb, lipsum}  
\usepackage[dvipsnames]{xcolor}
\usepackage[normalem]{ulem}
\newcommand{\BE}[0]{\begin{equation}}
\newcommand{\EE}[0]{\end{equation}}
\newcommand{\BSES}[0]{\begin{subequations}}
\newcommand{\ESES}[0]{\end{subequations}}
\newcommand{\BEA}[0]{\begin{eqnarray}}
\newcommand{\EEA}[0]{\end{eqnarray}}
\newcommand{\jbbl}[0]{J_2} 
\newcommand{\jbbu}[0]{J_2'} 
\newcommand{\jbv}[0]{J_1} 
\newcommand{\fr}[0]{f} 
\newcommand{\jzzcc}[0]{J_{zz}^c} 
\newcommand{\jstcc}[0]{J_{st}^c} 
\newcommand{\frca}[0]{f_{c_1}} 
\newcommand{\frcb}[0]{f_{c_2}} 
\newcommand{\phic}[0]{\varphi_c} 
\newcommand{\gfld}[0]{\mathfrak{g}} 
\newcommand{\const}[0]{\Omega} 
\newcommand{\fld}[0]{\phi} 
\newcommand{\dld}[0]{\theta} 
\newcommand{\opa}[0]{O_1} 
\newcommand{\opb}[0]{O_{2}^{q}} 
\newcommand{\opba}[0]{O_2^{\pi}} 
\newcommand{\opbb}[0]{O_2^{\pi/2}} 
\newcommand{\opc}[0]{O_3} 
\newcommand{\opd}[0]{O_4} 
\newcommand{\ope}[0]{O_5} 
\newcommand{\lpa}[0]{\lambda_1} 
\newcommand{\lpb}[0]{\lambda_{2}^{q}}  
\newcommand{\lpc}[0]{\lambda_3} 
\newcommand{\lpd}[0]{\lambda_4} 
\newcommand{\lpe}[0]{\lambda_5}

\newcommand{\hsat}{h_{\scalebox{0.7}{sat}}}
\newcommand{\rlm}{\rangle_{\scalebox{0.7}{lm}}}
\newcommand{\nmax}{n_{\scalebox{0.7}{max}}}
\newcommand{\hcr}{h_{\scalebox{0.7}{cr}}}
\begin{document}
\title{
  Enhancement of magnetization plateaus  in low dimensional spin systems
}  
\author{Alexandros Metavitsiadis}\email{a.metavitsiadis@tu-bs.de}
\affiliation{Institute for Theoretical Physics, Technical University 
  Braunschweig, D-38106 Braunschweig, Germany} 
\author{Christina Psaroudaki}\email{cpsaroud@uni-koeln.de}
\affiliation{Department of Physics, California Institute of Technology, 
Pasadena, CA 91125, USA} 
\affiliation{Institute for Theoretical Physics, University of Cologne, 
D-50937 Cologne, Germany} 
\author{Wolfram Brenig}\email{w.brenig@tu-bs.de} 
\affiliation{Institute for Theoretical Physics, Technical University 
  Braunschweig, D-38106 Braunschweig, Germany} 
\date{\today}

\begin{abstract}
We study the low-energy properties and, in particular, the magnetization process 
of a spin-1/2 Heisenberg $J_1-J_2$ sawtooth and frustrated chain (also known 
as zig-zag ladder) with a spatially anisotropic $g$-factor. We treat the problem both
analytically and numerically while keeping the $J_2/J_1$ ratio generic. Numerically, 
we use complete and Lanczos diagonalization as well as the infinite time-evolving block 
decimation (iTEBD) method. Analytically we employ (non-)Abelian 
bosonization.  Additionally for the sawtooth chain, we provide an analytical description 
in terms of flat bands and localized magnons. By considering a specific pattern for the
$g$-factor anisotropy for both models, we show that a small anisotropy significantly
enhances a magnetization plateau at half saturation.  For the magnetization of the 
frustrated chain, we show the destruction of the $1/3$ of the full saturation 
plateau in favor of the creation of a plateau at half-saturation. 
For large anisotropies, the existence of an additional plateau at zero magnetization  
is possible. Here and at higher magnetic fields, the system is locked in the 
half-saturation plateau,  never reaching full saturation.  
\end{abstract}

\maketitle

\section{Introduction} 
Frustrating interactions in quantum magnets have revealed a plethora of 
exotic phenomena with no classical analogue \cite{Balents2010, Starykh_2015}. 
One such example is the appearance of magnetization plateaus, i.e., 
regions in the magnetization process of a paramagnetic system at which the
magnetization stays put at some fractional 
value $M_p$ of the saturation magnetization $M_s$ despite the increase of the 
magnetic field.  Magnetization plateaus have been observed in  several systems 
independent of their dimensionality,  
described by very different geometries, e.g. in Shastry-Sutherland type 
of models \cite{SRIRAMSHASTRY19811069, PhysRevLett.111.137204, 
PhysRevB.90.104404}, 
triangular \cite{Chubukov_1991, PhysRevB.75.134412, 
s41467-018-04914-1, Ono_2004},  square \cite{doi:10.7566/JPSJ.85.094708,
 PhysRevLett.85.3269}, checkerboard \cite{PhysRevB.94.140404}, 
Kagome geometries \cite{doi:10.7566/JPSJ.84.114703},   
down to one-dimensional (1D) frustrated systems \cite{doi:10.1143/JPSJ.74.1849, 
doi:10.1143/JPSJ.66.1900} and many more (see also 
Refs.~\cite{Takigawa2011, Honecker_1999, Honecker_2004, doi:10.1063/1.2008130,
 PhysRevLett.88.167207} for comparative studies).  
 While significant knowledge may have been gathered on the ground state 
 of these systems, the situation often becomes more challenging at the 
 magnetization plateau where a prerequisite for the existence of a plateau is the 
opening of a gap in some parts of the spectrum. 

Here we use the sawtooth as well as the frustrated chain  
(see Fig.~\ref{fig:model}) as  prototypical models to investigate the effect of a spatially
 modulated $g$-factor in systems that exhibit magnetization plateaus. 
Not only are these models the cornerstones of one-dimensional quantum magnetism 
but they have also been used to understand physics in higher dimensions. 
According to the Oshikawa-Yamanaka-Affleck  theorem 
\cite{PhysRevLett.78.1984, PhysRevLett.79.5126},  
a one-dimensional spin-$S$ system with a $p$-periodic ground state, 
could exhibit magnetization plateaus for values of the magnetization $M$ 
which satisfy the condition $ pS(1-M/M_s) \in \mathbb{Z}$. 
The sawtooth chain exhibits a magnetization plateau 
at half saturation $M_p = M_s/2$ for a wide range of the ratio 
$\jbbl/\jbv$ \cite{doi:10.1142/S0217979208050176}, 
while the frustrated chain  exhibits  a magnetization plateau at $M_s/3$ 
\cite{PhysRevB.69.174409}.  

In this work, we primarily focus on the sawtooth chain, but we keep the analysis 
as general as possible to simultaneously treat the frustrated chain  
and discuss the similarities as well as the differences between the two models.  
Although the sawtooth chain, as well as variants of it, have been 
studied theoretically early on \cite{doi:10.1143/JPSJ.57.1891, PhysRevB.48.10552,
MONTI1991197, doi:10.1143/JPSJ.64.695, PhysRevB.51.305, PhysRevB.53.6393, 
PhysRevB.53.6401, NAKAMURA1997315,  PhysRevLett.81.445, PhysRevLett.87.087205, 
refId2, Richter_2004,  doi:10.1143/JPSJ.75.114712, doi:10.1143/JPSJ.77.044707,  
PhysRevB.77.064413}  they remain of great interest until today \cite{ refId1, 
PhysRevB.84.094452, PhysRevA.87.013607, JIANG201530, Dmitriev_2018, Paul2019}.
From an experimental point of view,  the situation remains challenging, with 
only a limited number of compounds being  reported until this day to materialize 
dominant magnetic 
interactions in a sawtooth pattern. Prominent examples are the delafossite 
$\mathrm{YCuO_{2.5}}$  \cite{CAVA1993437, Capponi_2007, VANTENDELOO2001428,
 PhysRevB.71.014432}, the double spin chain systems $\mathrm{K Cu Cl_3}$  
and $\mathrm{Tl Cu Cl_3}$ \cite{doi:10.1143/JPSJ.66.1900}, the multiferroic 
$\mathrm{Mn_2 Ge O_4}$ \cite{Honda2017, doi:10.1143/JPSJ.81.103703,  
PhysRevLett.108.077204}, the olivines $\mathrm{Zn}{L}_{2}{\mathrm{S}}_{4}$ $
(L=\mathrm{Er},\mathrm{Tm},\mathrm{Yb})$ \cite{PhysRevB.73.012413}, and  
the Fe chains $\mathrm{Rb_{2} Fe_{2} O (AsO_{4})_{2}}$ and 
${\mathrm{Fe}}_{2}\mathrm{O}({\mathrm{SeO}}_{3}{)}_{2}$ 
\cite{PhysRevB.89.014426, PhysRevB.99.064413}. Remarkably, and despite great efforts, 
 a magnetization plateau 
has not been reported for any of these systems. Very recently, 
a study on the magnetic structure of the natural mineral atacamite showed that its  
magnetic structure is that of the sawtooth type with aniferromagnetic couplings between 
the  spin-1/2 moments, with however a puzzling magnetization plateau
\cite{heinze2019atacamite}.  

Our paper is organized as follows. First, in Sec.~\ref{sec:model}, we present the 
model and its basic properties. In the first part of our results,  we present 
analytical calculations (Sec.~\ref{sec:analytical}), first in terms of field 
theory,  Secs.~\ref{sec:fieldtheoryn} and \ref{sec:fieldtheorya},  and then 
in terms of localized magnons in Sec.~\ref{sec:localized}.  In the second part
of our analysis, we present numerical results (Sec.~\ref{sec:numerical})  for 
a uniform or modulated $g$-factor and for both the sawtooth as well as the 
frustrated chain. We conclude in Sec.~\ref{sec:conclusions}.  
\section{Model}\label{sec:model} 
\begin{figure}[t!] 
  \centering 
  \includegraphics[width=0.90\columnwidth]{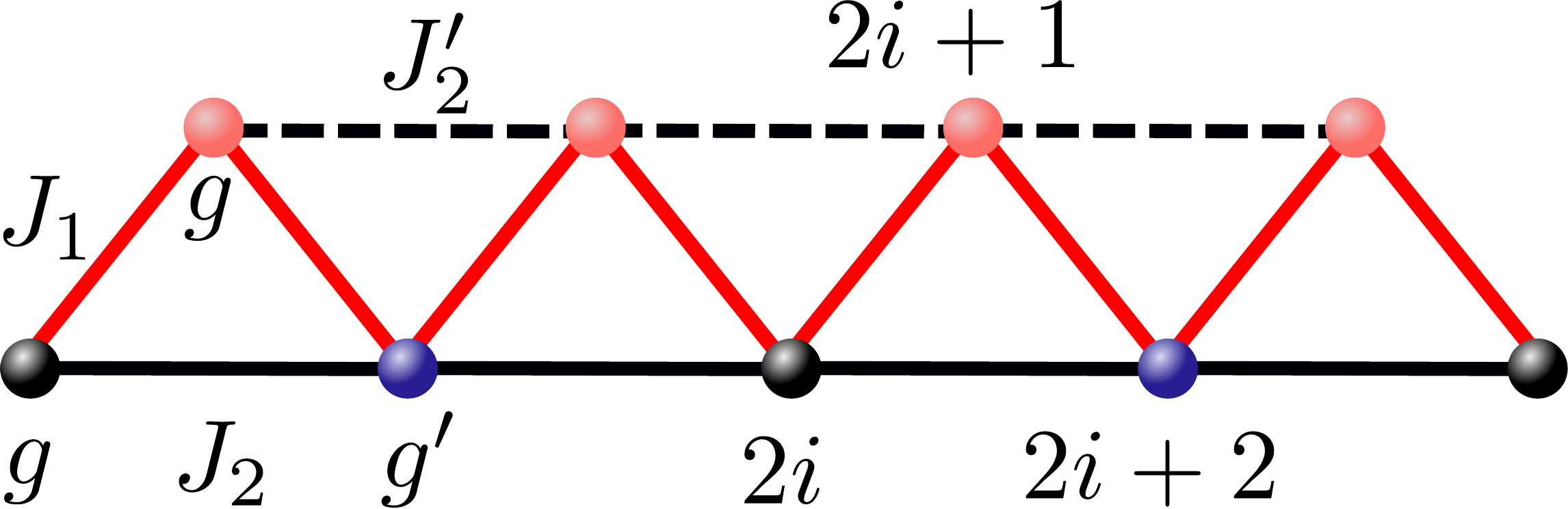} 
  \caption{A generalized 1D chain with anisotropic next nearest neighbor 
  interactions. The upper base-base coupling is parametrized as 
  $\jbbu = (1- \alpha) \jbbl$, with $\alpha =0,1$. 
  For $\alpha=0$ ($\jbbu=\jbbl$) the frustrated chain with NNN interactions, 
  or zig-zag ladder,  is recovered while for $\alpha=1$ ($\jbbu=0$) the sawtooth 
  chain is recovered. The $g$-factor is considered to be either uniform 
  $g'= g$ or to vary on every other site of the lower chain with $g' = g -\delta g$ 
  and $\delta g>0$, indicated by black and blue colors. For the upper chain, we 
  consider a uniform $g$ value. 
}\label{fig:model} 
\end{figure}
Our starting point is the generalized 1D Heisenberg chain with nearest 
$\jbv$ neighbor (NN) and anisotropic next nearest neighbor (NNN) interactions 
$\jbbl$ and $\jbbu =(1-\alpha)\jbbl $, Fig.~\ref{fig:model}.  
The Hamiltonian that describes this system in the  presence of a uniform 
magnetic field along the $z$-axis
$\mathbf{B}=B \hat z$ reads 
\BEA
H &=&   - \sum_j  g_j \mu_B B S_j^z + \jbv \sum_j \mathbf{S}_{j} \cdot  \mathbf{S}_{j+1} \nonumber \\ 
&+&
 \frac{\jbbl}{1+\alpha} \sum_j [1+(-1)^j\alpha] \mathbf{S}_{j} \cdot  \mathbf{S}_{j+2} ~. 
\label{eq:spinHamiltonian}   
\EEA
where $\mathbf{S}_{j}$ are spin-$\frac{1}{2}$ operators residing on the lattice  
sites, $\mu_B$ is the Bohr magneton, and we set $\hbar =1$. We consider 
only two values for the parameter $\alpha=0, \textrm{ or } 1$. 
The sawtooth chain is recovered for $\alpha=1$ ($\jbbu=0$) and the 
frustrated chain (or zig-zag ladder) for $\alpha=0$ ($\jbbu = \jbbl $) respectively. 
Although we are mainly interested in the case of the sawtooth chain, 
we keep $\alpha$ as a parameter in our analysis to draw analogies between the two models. 
The ratio of the two couplings $\fr = \jbbl/\jbv$ can also be perceived as the degree of 
frustration. 
A central point of this work is our consideration of a particular spatial variation 
of the $g$-factor. Namely,  $g_j$ exhibits  
two patterns: a \emph{uniform} one $g_j = g$ and a \emph{modulated} 
one, where the value of the $g$-factor on every second site of the lower chain has 
a different value $g'$, with $g' = g - \delta g$ and $\delta g > 0 $. 
In most material realizations, a possible finite $\delta g$ is expected to be 
of the order $\delta g/g \sim \mathcal{O}(0.1)$. Despite that, here,   
we vary $\delta g$  as a free parameter, letting $g$ to acquire values 
as high as $g$, to provide a complete picture of our theoretical findings. 
Note that for $\delta g > g$, the $g$-factor exhibits a staggering sign. 
We would also like to stress that the sawtooth chain has no leg inversion symmetry, 
and therefore such a modulation only on one part of the system is not unlike to 
happen in material realizations. 

The two models, the frustrated chain and the sawtooth chain share some common 
properties. They both exhibit either a unique gapless spin fluid (Tomonaga 
Luttinger liquid) ground state, or a  gapped dimerized one when the 
degree of frustration is in the range $\frca < \fr < \frcb$ \cite{PhysRevB.54.9862, 
PhysRevB.54.R9612, refId2, PhysRevB.81.104406}. 
Both models allow for analytical solutions of their ground states at special values 
of the frustration ratio ($\fr=1$ for the sawtooth chain 
and $\fr=1/2$ for the frustrated chain) with double degenerate ground states 
\cite{doi:10.1063/1.1664978, doi:10.1063/1.1664979, CASPERS1982103,
 MONTI1991197}. The low lying excitations are kink and anti-kinks in the form 
 of domain walls spatially separating 
regions of one type of ground state. Their dispersion, 
however, differs with the kink excitations being gapped in the sawtooth  
and gapless in the frustrated chain  \cite{PhysRevB.53.6401}.  
Another difference between the two models appears 
in the magnetization process of each system. While the sawtooth chain exhibits a 
plateau at $M_p=M_s/2$, the frustrated chain exhibits one at $M_p = M_s/3$. 
For completeness, we mention that the value of the plateau $M_p$ depends on the 
geometrical properties of the model and therefore is independent of the 
coupling ratio $\fr$ in contrast to the plateau's width, 
which depends on the size of the gap in the presence of the magnetic field,  
and therefore depends on the degree of frustration. 
\section{Analytical results}\label{sec:analytical}
First, we  treat the problem analytically by employing Abelian and 
non-Abelian bosonization focusing on the $\jbv \gg \jbbl, \jbbu$ regime.
\subsection{Non-Abelian bosonization} \label{sec:fieldtheoryn}
Let us first detour by revisiting the field theory of the sawtooth chain 
in the absence of a magnetic field in the context of non-Abelian bosonization 
\cite{book-cft, affleckleshouches, PhysRevB.72.094416,book-senechal-2004,tsvelik}. 
Within non-Abelian bosonization, both the $U(1)$ and the $SU(2)$ symmetries of 
the underlying Hubbard model are considered in terms of the bosonic field 
$\phic$ and the matrix field $\gfld$. The charge sector is gapped out,  
and the spin operators can be written in terms of  chiral $SU(2)$ currents 
$\mathbf{J}_{L/R}$ and the  staggered magnetization 
$\mathbf{n} = \mathrm{Tr} \bm{\sigma} \gfld$ as 
\begin{equation} \label{eq:SpinOperator}
\mathbf{S}(x)\approx
\mathbf{J}_L(x)+\mathbf{J}_R(x) + (-1)^{x}~\const~\mathbf{n}(x)~,  
\end{equation}
where the bosonization constant $\const$ is of the order of one, and it is 
related to the mass of the charge sector. The field theory is completed by 
considering  one more additional operator, the dimerization $\epsilon$, 
given by the non-oscillating part of 
$ \epsilon(x) \sim  (-1)^x \mathbf{S}(x) \cdot \mathbf{S}(x+a) \sim \mathrm{Tr} (\gfld)$   
\cite{PhysRevB.72.094416, PhysRevB.95.144415}.     

In the $\jbv \gg \jbbl, \jbbu$ regime the system can be considered as a 
Heisenberg chain with a coupling constant $\jbv$ perturbed by the couplings 
$\jbbl, \jbbu$, where each one of the latter couples  NNN sites that belong 
only to one of the two sublattices (the upper or the lower chain). 
In the continuum, the perturbation of the fixed point Hamiltonian $H_0(\jbv)$ reads  
\BE
\delta H  = \frac{1}{1+\alpha}  
\int dx \left[\lambda_{J} \mathbf{J}_L \cdot  \mathbf{J}_R (x) +  \lambda_{\partial \epsilon} 
\partial \epsilon(x)    \right] ~. 
\EE
The bare couplings $\lambda$ depend on the microscopic couplings $\jbv, \jbbl, \jbbu$ 
and the bosonization constant $\const$
$$ 
\lambda_J \sim J^c - \jbbl ~, \quad 
\lambda_{\partial \epsilon} \sim \alpha \frac{3 \const^2}{2\pi}    \jbbl ~, 
$$ 
with $J^c$ the critical coupling for each model. 
The current operator is generated by 
the interaction term of the NN Hamiltonian as well, $\sim \sum S_j^z S_{j+1}^z$, 
and the NNN couplings modify its bare value. 
For the frustrated chain ($\alpha=0$), translation symmetry by one site 
is restored, and the Luttinger liquid fixed point is solely disturbed by the current 
operator, which is known to open a gap at 
$\jzzcc / \jbv \approx 0.241167$ and drive the system in a dimerized phase \cite{PhysRevB.54.R9612}.   
On the other hand, for the  sawtooth chain ($\alpha=1$), the strength of 
the current operator due to the NNN interactions is reduced by a factor of 1/2  while 
the $\partial \epsilon$ operator appears. The effect of this operator, which is a total
 derivative,  
has triggered a big dispute in the literature \cite{PhysRevB.65.172408, 
PhysRevLett.87.087205,PhysRevB.67.054412,  PhysRevLett.89.149701, 
PhysRevLett.89.149702}. Leaving aside for a moment the $\partial \epsilon$ operator, 
the operator contents of the two models are identical and 
the only difference arises in the bare coupling $\lambda_J$.  
This means that the sawtooth chain would undergoe a phase transition to a gapped 
dimerized phase at $\jstcc = 2\jzzcc \approx 0.48 $, which  coincides remarkably 
with the value
predicted from numerical simulations \cite{refId2}. In retrospect, one can argue that 
the $\partial \epsilon$ operator has no effect on the deformation of the 
critical lines  and can, therefore, safely be ignored. 
\subsection{Abelian bosonization} \label{sec:fieldtheorya}
\par 
Next, we move to the case of interest, i.e., the sawtooth chain in the presence 
of a magnetic field with an anisotropic $g$-factor. Because of the presence of the 
magnetic field, $\mathrm{SU(2)}$ symmetry is broken, and we turn to Abelian bosonization 
\cite{tsvelik, book-senechal-2004, book-giamarchi}. 
The low energy properties of these systems in the presence of a uniform magnetic 
field have been  described extensively in the literature 
\cite{PhysRevLett.87.087205, PhysRevB.67.054412, Sarkar_2005, PhysRevB.65.172408, PhysRevB.77.064413, PhysRevB.57.3454}. Here, 
we  include only what is essential for our work. 

In the standard Abelian Bosonization machinery, spin operators are described in terms 
of fermionic operators using a  
Jordan-Wigner transformation.  The spectrum of the XY-NN-Hamiltonian is linearized 
around the Fermi points $\pm k_F$ and slow varying chiral field operators are introduced 
in the continuum ($x=ja$)
\BE\label{eq:RL} 
\psi (x) \sim e^{ik_F x} \psi_R(x) + e^{-ik_F x} \psi_L(x)~. 
\EE 
The Fermi wavevector is given in terms of the magnetization 
$M$ and the saturation magnetization $M_s = L/2$,  
$k_F = \frac{\pi}{2a}(1-m)$, with $m=\frac{M}{M_s}$; 
for example $k_F=\frac{\pi}{2a}$ for $M=0$ 
and $k_F=\frac{\pi}{4a}$ for $M=M_s/2$. In turn, the $\psi_{L,R}$ fields are expressed 
in terms of the $U(1)$ bosonic fields $\fld = \fld_R+\fld_L$ according to 
\BE 
\psi_R(x) = \frac{1}{\sqrt{2\pi a}} e^{-i \beta \fld_R (x)}~, ~~ 
\psi_L(x) = \frac{1}{\sqrt{2\pi a}} e^{+i \beta \fld_L (x)}  ~, 
\EE 
with $\beta$ a numerical constant, here $\beta = \sqrt{4\pi}$. 
The system is then described in terms of $\fld$ and its dual field 
$\dld = \fld_R-\fld_L$ with $[\fld(x),\dld(x')] = -i\vartheta(x-x')$ and 
$\vartheta$ the Heaviside step function.  

Here we assume a spatially modulated $g$-factor, which exhibits an 
alternating pattern taking values $g$ or $g'= g- \delta g$ on the lower chain,
as described in Fig.~\ref{fig:model}.  The effect of this modulation is that some 
of the spins, the ones residing on the black sites of Fig.~\ref{fig:model}, 
experience an effective magnetic field which is reduced from the uniform 
value $h=g \mu_B B $, to a different value $h'=g'\mu_B B$. The site dependence 
of the effective magnetic field $h(x)$ can be written as 
\BE
h(x)  = h - \frac{\delta h}{4}  \left[1 + 
2\cos \left(\frac{\pi}{2a} x\right) + \cos \left(\frac{\pi}{a} x\right)  \right]~, 
\EE
with $\delta h = \delta g \mu_B B$. Therefore it becomes apparent that 
in Fourier space, $h(x) = \sum_q h_q e^{iqx}$,  the 
effective magnetic field has a finite overlap not only at momentum $q=0$, 
but at $q = \pm \frac{\pi}{2a} \textrm{ and } \pm \frac{\pi}{a}$ as well, with the corresponding Fourier 
components $h_{q} = h- \frac{\delta h}{4}, \frac{\delta h}{4} \textrm{, and } 
\frac{\delta h}{8}$.  

In the field theory representation,  the system is 
described by the Tomonaga-Luttinger liquid (TLL) Hamiltonian $H_0$,  
perturbed by several operators 
\BEA
H &=& H_0 +\sum_{j}  \int dx   \lambda_j O_j(x), ~\textrm{with, }\\ 
H_0 &=& \frac{v}{2} \int dx \left[K[\partial \dld(x)]^2 + \frac{1}{K}[\partial \fld(x)]^2\right] \nonumber ~, 
\EEA 
and  $v=v(m,\jbv,\jbbl)$, $K=K(m,\jbv,\jbbl)$ the Tomonaga-Luttinger (TL) 
parameters \cite{PhysRevB.65.172408, PhysRevB.57.3454}. The perturbative part 
of the Hamiltonian reads  
\BSES\label{eq:operatorContent}
\BEA
 \opa &=& \partial \fld,~  
 \opb  =  \cos [\beta \fld(x)  -(2k_F-q) x], \label{eq:opaopb} \\
 \opc &=&  \cos\left[2\beta \fld(x) - (4 k_F -G)x -2 k_F a  \right], \label{eq:opc}\\ 
 \opd &=&   \cos\left[2\beta \fld(x) - (4 k_F -G)x -4 k_F a  \right], \label{eq:opd}  \\ 
 \ope &=&  \cos\left[2\beta \fld(x) - (4 k_F -\pi/a)x -4 k_F a  \right], \label{eq:ope} 
\EEA 
\ESES
with 
\BSES
\BEA\label{eq:operatorCouplings}
 \lpa &=& -\frac{h_0}{\sqrt{\pi}}   + C(\alpha,m,J_1,J_2),      \quad 
 \lpb =  -\frac{h_q}{\pi a},~ \\
 \lpc &=&  \frac{\jbv}{2 \pi^2 a},~~
 \lpd \sim  \frac{ \jbbl}{2\pi^2a (1+\alpha)},~~
 \lpe = \alpha \lpd~. 
\EEA 
\ESES
\newcounter{comments}
Several comments are in order regarding  this  rich operator content.  
(\addtocounter{comments}{1}\roman{comments}) $G$ is the reciprocal lattice vector 
which for a uniform chain reads $G=\frac{2\pi}{a}$.  For a system with a 
unit cell involving $\nu \ge 1 $ sites,  it is modified to $G = \frac{2\pi}{\nu a}$.  
(\addtocounter{comments}{1}\roman{comments}) The operators 
in Eq.~\eqref{eq:opaopb} arise due to the magnetic field  whereas 
the rest from the Heisenberg interactions.  
(\addtocounter{comments}{1}\roman{comments}) The $\opb$ operator 
depends on $q$ due to the two Fourier components at $q=\frac{\pi}{2a}$ 
and  $\frac{\pi}{a}$ of the effective magnetic field.  
Furthermore, there are two different $\lpb$ for each $q$ component. 
(\addtocounter{comments}{1}\roman{comments}) Not all of these operators survive 
at every magnetization and/or for any $G$.  The rapidly oscillating factors in the
arguments of the cosines make  them vanish under integration, unless the terms in the 
parentheses multiplying $x$ vanish.
(\addtocounter{comments}{1}\roman{comments}) Four-fermion operators   
yield in the continuum operators that oscillate with a momentum $q=2k_F$ as well. 
Normally these terms oscillate and will vanish upon integration. However, for 
a non-vanishing $\alpha$ their momentum dependence is modified to  
$q=2k_F-\pi/a$, which vanishes  for  $k_F=\frac{\pi}{2a}$ ($m=0$). 
One needs to be careful, since they could yield relevant operators 
\cite{PhysRevB.57.5812,PhysRevLett.87.087205}, however, 
at the plateau, $k_F = \frac{\pi}{4a}$,  they oscillate and we drop them 
here for simplicity. 
(\addtocounter{comments}{1}\roman{comments}) 
At finite magnetization, namely away from half filling in the fermion representation, 
there are additional contributions to $\lpa$ due to the finite chemical potential. 
These terms are incorporated in the constant $C$. 
(\addtocounter{comments}{1}\roman{comments}) 
Interactions modify the contribution of the  
next nearest neighbor umklapp terms, which arise also from their $XY$ part of 
the spin Hamiltonian. Far from the non-interacting regime,  
the exact coefficients of the umklapp terms  cannot be accurately established.  

The operators in Eq.~\eqref{eq:opaopb} is more relevant than  the cosine operators  
in Eqs.~\eqref{eq:opc}-\eqref{eq:ope}, which can  be marginal, relevant, or irrelevant. 
This depends on the coefficient of $\fld$ in their argument as well as on the TLL  
interaction parameter $K$. To determine the behavior of the cosine operators 
one needs to consider its  behavior under the renormalization group where the 
momentum cut-off $\Lambda $ is decreased according to 
$\delta \Lambda(l)  = -\Lambda(l)\delta l $. Assuming the sine-Gordon 
 Hamiltonian  $H=H_0 + g\int dx \cos[ \gamma \fld(x)]$ the coupling $g$ of the 
 operator $\cos \gamma \fld $ changes in first order according to 
$\frac{\delta \ln g}{\delta \ln \Lambda} = d_\gamma -2$, with 
$d_\gamma=\frac{K \gamma^2 }{4 \pi}$ 
the scaling dimension of this operator. This means that  the cosine operator 
is relevant for $d_\gamma<2$, marginal for $d_\gamma = 2$, and irrelevant for 
$d_\gamma >2$ \cite{book-senechal-2004,PhysRevB.57.5812}. 
For example, for vanishing $\jbbl=0$ where $K=1/2$, 
the coefficient in the umklapp term in Eq.~\eqref{eq:opc} is 
$\gamma = 2\beta = 2\sqrt{4\pi}$, i.e., $d_{2\beta} =2$,   
namely the operator is marginal which agrees with the literature \cite{book-giamarchi}. 

Let us now discuss the effect of the magnetic field combined with the site 
modulation of the $g$-factor. The magnetic field contribution is described in 
Eq.~\eqref{eq:opaopb}. From there, we see that the strength of the 
$\partial \fld $ operator is reduced from its uniform value $h$ by $\delta h/4$. 
This operator tends to make the field fluctuate, preventing 
its pinning to some constant value, which in turn would open a gap, 
and the formation of a magnetization plateau would become possible.  
In other words, $\partial \fld $ is responsible for destroying magnetization plateaus, 
and in the presence of the modulation is weakened.  
The second contribution of the magnetic field comes in the form of the 
cosine operator in the same equation where the spatial modulation of the 
$g$-factor yields the $\cos \beta \fld$ term, which is always relevant since 
$K<2$ \cite{Sarkar_2005}. Therefore, since this operator is more relevant than the 
rest of the operators, it is highly probable to prevail under renormalization 
and drive the system in a gapped phase, when it is not oscillating. 
From the above, it becomes apparent that the $g$-factor modulation has a 
twofold effect on the interaction of the sawtooth with the magnetic field. 
First, it reduces the strength of the operator destabilizing magnetization plateaus,
and second, it yields new relevant operators that can stabilize magnetization 
plateaus. 

\subsubsection*{Sawtooth chain}

We now apply  the above to the sawtooth chain  ($\alpha=1$). 
For a uniform magnetic field ($\delta h = 0 $), 
the lattice periodicity is determined by the two-site unit cell of the  
microscopic model and, therefore $G=\pi/a$. The physics at zero magnetization 
has been described in terms of non-Abelian bosonization in Sec.~\ref{sec:analytical}.  
As the uniform magnetic field increases, the $\partial \fld $ operator,  
which can be absorbed by the  substitution $\fld \rightarrow \fld + \frac{\lpa K}{v} x$,  
drives the system to an incommensurate phase still described by a TLL fixed point. 
At some point, a gap opens due to the operators in Eqs.~\eqref{eq:opc}-\eqref{eq:ope}, 
and the system enters the plateau phase. This happens at 
$m = 1/2$ where $4 k_F = G= \frac{\pi}{a}$ and the oscillating  factors 
in the argument of the cosines vanish \cite{Sarkar_2005, PhysRevB.65.172408}.   

By introducing the $g$-factor modulation, the reciprocal lattice vector becomes 
$G=\frac{\pi}{2a}$ and the operators $\opc$ and $\opd$ do not contribute. 
Hence, there is a competition between the $\ope$ and the more relevant $\opbb$ 
($d_\beta < d_{2\beta}$) 
which is expected under RG to reach first the strong coupling limit. 
The gap of the system scales as $\Delta \sim e^{-\ell} $ where $\ell$ is the 
point where the perturbative RG breaks down, and the system is no 
longer conformally invariant \cite{book-giamarchi}. 
Therefore, since more relevant operators reach  the strong coupling limit faster, 
one could expect a larger gap, meaning a broader magnetization plateau at $m=1/2$. 
This can also be understood from the sine-Gordon model, where
the soliton mass relates directly to the plateau width \cite{PhysRevB.57.3454} 
and the soliton mass scales inversely proportionally to the argument of 
the cosine,  at least to leading order in $\gamma$. 
As a side remark, we mention that the operator $\opba$  survives at 
$k_F = \frac{\pi}{2a}$, albeit with a reduced bare coupling as compared to the coupling 
of $\opbb$,  and therefore with fine-tuning of the microscopic parameters 
a magnetization plateau at $m=0$ may, in principle, arise. 

\subsubsection*{Frustrated chain} 

The frustrated chain ($\alpha=0$) at a uniform field is known to exhibit a magnetization
plateau at $m=1/3$ due to an even less relevant umklapp operator 
\cite{PhysRevB.69.174409,doi:10.1143/JPSJ.74.1849}. 
However, when the site modulation dependence of the $g$-factor is switched on, 
the periodicity of the model changes, now $G=\frac{\pi}{2a}$, and the more
relevant operator $\opbb$ will be present and easily prevail. In fact, the rest of the 
operators in Eq.~\eqref{eq:operatorContent} vanish due to the oscillating factors 
in the argument of the cosine, and it can be safely assumed that the TLL fixed point 
is solely perturbed  by the $\opbb$ operator. Therefore, a wide plateau is expected at 
$m=1/2$ instead of $m=1/3$.   

\subsection{Localized magnons}\label{sec:localized}

\begin{figure}[t!] 
  \centering 
  \includegraphics[width=0.90\columnwidth]{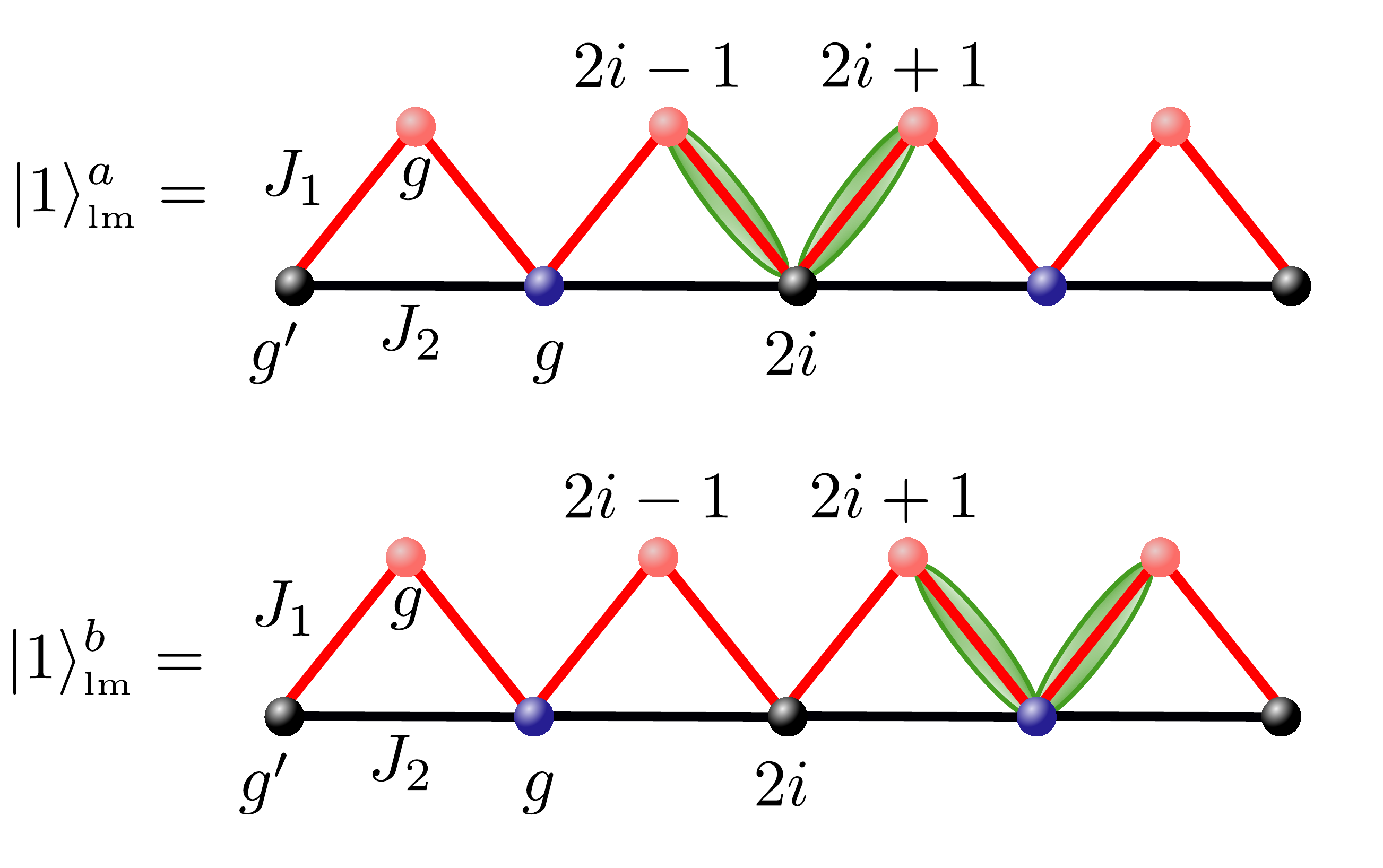} 
  \caption{Localized magnon states realized in the Heisenberg sawtooth chain. 
  The magnon lives on the restricted area indicated by the green ellipses. 
  State $\vert 1 \rangle_{\mbox{\tiny lm}}^a$ 
  with energy $\varepsilon_1^a = h-4 J_2 -(2/3)\delta h$ is the lowest eigenstate 
  in the sector $M = L/2-1$, while state $\vert 1 \rangle_{\mbox{\tiny lm}}^b$ is a 
  state with higher energy $\varepsilon_1^{b} =h-4 J_2$. 
}\label{fig:LocalMagn} 
\end{figure}

We now discuss how the magnetization plateau of the sawtooth chain can be explained 
in terms of localized magnons that emerge from the frustrated Hamiltonian of 
Eq.~\eqref{eq:spinHamiltonian} due to a flat energy dispersion relation.
Flat dispersions exist for several  strongly frustrated spin lattices 
\cite{doi:10.1063/1.2008130, doi:10.1142/S0217979215300078},  including 
the 2D Kagom\'{e} lattice \cite{PhysRevLett.88.167207} and 3D pyrochlore 
lattice \cite{Richter_2004}, and frustrated electronic systems 
\cite{PhysRevLett.99.070401,PhysRevB.78.125104}.   

We first note that in the subspace $M = M_s = L/2$, 
with $L$ being the total number of spins,  the fully polarized state $\vert FM \rangle $ 
becomes the ground state for sufficiently large magnetic fields exceeding the saturation 
field $\hsat$, and plays the role of the vacuum state,  $|0\rangle = \vert FM \rangle$,  
for the magnon excitations. 
For $\delta g=0$ the one magnon state reads
\BE 
\vert 1 \rangle_{k}=\sum_{i=0}^1 \frac{1}{N_i} \sum_{j=1}^{L/2} e^{i2kj} S_{2j+i}^{-} 
\vert 0 \rangle~,
\EE  
where  $S^-=S^x-iS^y$,  $N_i$ normalization constants, and 
$k= 4\pi \frac{l}{L}$,  with $l \in \mathbb{Z}$ in the range $[0,\frac{L}{2})$. For $f=1/2$,  it 
corresponds to a completely flat magnon band 
$\varepsilon_1 = h - 4J_2$ \cite{doi:10.1142/S0217979208050176}.  
 A complete flat dispersion suggests that one can construct a localized magnon state 
 in a finite region of the lattice of the form 
\BE
 \vert 1 \rlm  =l_{2j}^{\dagger}\vert 0 \rangle=
 \frac{1}{\sqrt{6}}(S_{2j-1}^{-} -2 S_{2j}^{-} +S_{2j+1}^-)\vert 0 \rangle,
 \label{eq:LocMag}
\EE 
 where the magnon is trapped in a valley indicated by the green ellipses in
  Fig.~\ref{fig:LocalMagn}. 
  Under general assumptions, one can demonstrate that $\vert 1 \rlm$ is the lowest 
   eigenstate 
  in the sector $M = M_s-1$, and becomes the ground state in an appropriate magnetic 
   field 
  \cite{Schnack2001,Schmidt_2006}.  
  
  Due to the localized nature of state $\vert 1 \rlm$, we proceed to fill the remaining of 
   the lattice with 
  $n$ localized magnons $\vert n \rlm = l_{2j}^{\dagger}\dots l_{2j'}^{\dagger}  
  \vert 0 \rangle$, 
  states of lowest energy in the sector $M = M_s-n$, with energy 
  $\varepsilon_n = n \varepsilon_1 = n(h-4 J_2)$ above the energy of the 
  ferromagnetic state. 
  In order to avoid magnon-magnon interactions, magnons are 
  constructed with sufficiently large space separation between them, 
  and $n$ cannot exceed $\nmax = L/4$. 
  We now allow for a finite but small $\delta g >0$. Although states $\vert n \rlm$ 
  are no longer eigenstates of the Hamiltonian, we can consider $\delta g \ll g$, 
  and calculate their energy within first order perturbation theory. 
  Two types of localized states can be realized, depending on whether
   the valley area is 
  centered around a site with $g'$ (black) or with $g$ (blue) 
  (see Fig.~\ref{fig:LocalMagn}). 
  After a straightforward calculation we find that states $\vert n \rlm^a$, 
  centered around a site with $g'$, have the lowest energy with 
  $\varepsilon_n^{a} = n(h-4 J_2 -(2/3)\delta h)$, 
  while states $\vert n \rlm^b$, centered around a site with $g$, 
  remain unaffected by $\delta g$ and have energy equal to 
  $\varepsilon_n^{b} = n(h-4 J_1)$. 
  Thus, states $\vert n \rlm^a$ are the lowest energy states 
  in the corresponding sector of magnetization $M$.  
  
Under the assumptions specified above, at the saturation field,  
\BE \label{eq:hsat}
\hsat =  \frac{12 g }{3g-2\delta g} \jbbl  ~,  
\EE 
 there is a complete degeneracy of all localized-magnon  states with energy 
 $\varepsilon_n =0$. As a result, $m$ jumps between  the saturation 
 value $m=1$ and the value $m=1-\nmax/M_s=1/2$, with $\nmax=L/4$. 
 This is a macroscopic quantum effect,  and the value of the jump vanishes if 
 the spins become classical. The result above shows that a finite $\delta g$ 
 shifts the saturation field towards larger values, corroborating the field theory 
 prediction for a larger plateau if  $\delta g \neq 0$. For higher values of the 
 anisotropy  $\delta g$, first-order perturbation theory 
is expected to fail. The full treatment of the problem is involved and is done 
by means of second-order perturbation theory, taking into consideration 
the overlap of localized states and propagating states with 
energy higher than $\varepsilon_n$. However, this is beyond the scope 
of this work, and we leave it as a motivation for future studies. 

%
%
%
\section{Numerical results}\label{sec:numerical}  
\begin{figure*}[t] 
\centering 
\includegraphics[width=2\columnwidth]{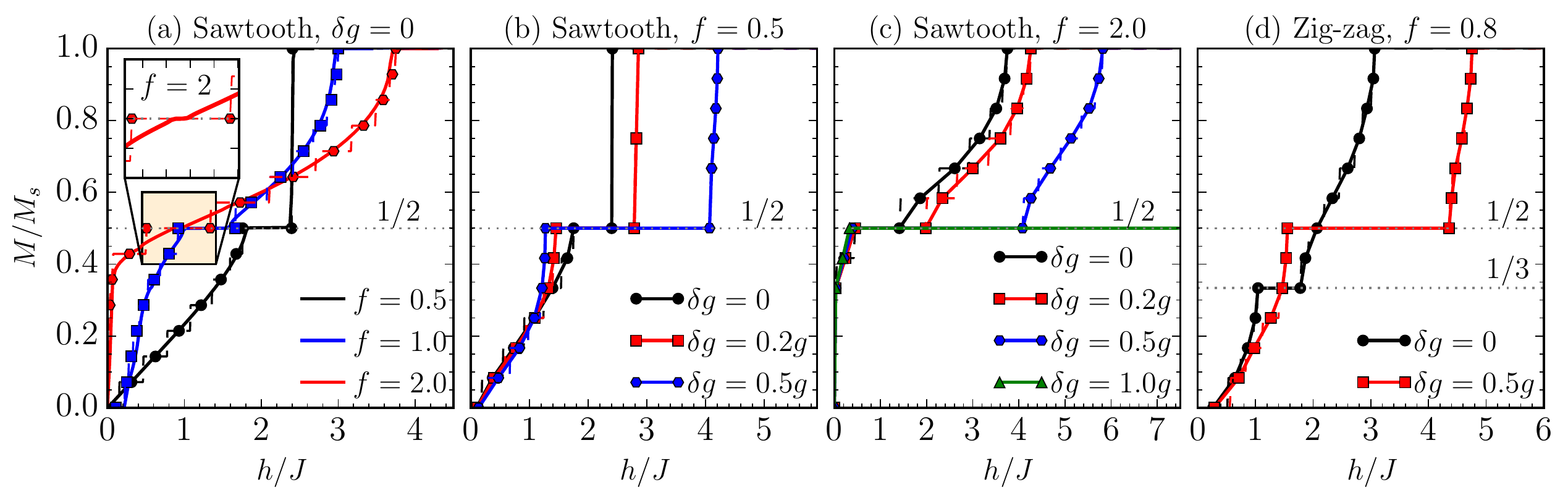} 
\caption{(color online) Magnetization of the sawtooth (a-c) and the frustrated (d) 
chain versus the magnetic field. The magnetic field axis is rescaled by 
$J=(2J_1+J_2)/3$  for the sawtooth and by $J=(J_1+J_2)/2$ for the frustrated chain.  
(a) Comparison of the sawtooth magnetization obtained via ED for $L=28$ sites 
(thin dashed lines and points) to that obtained via iTEBD (thick solid lines) for three 
values of the frustration parameter $\fr = 0.5,1,2$ and a uniform $g$-factor, $\delta g = 0$. 
The inset zooms in the highlighted region for $f=2$.   
 (b)  Sawtooth magnetization via ED for $L=24$ and $\fr=0.5$ for three different values 
of the $g$-factor modulation, $\delta g = 0,0.2g, 0.5g$. (c) Sawtooth chain's
 magnetization via ED
for $L=24$ and $\fr=2$ for four different values of the $g$-factor modulation, 
$\delta g = 0,0.2g, 0.5g, g$. (d) Frustrated chain's  magnetization via ED for $L=24$ 
and $\fr=0.8$ for two different values of the $g$-factor modulation, $\delta g = 0,0.5g$. 
The points in each panel and for each curve mark the middle of each magnetization step, 
except when a plateau is expected, 
where these points mark the beginning and the end of this step.}\label{fig:magnetization}  
\end{figure*}
To test the previous theoretical findings, we resort to numerics. 
We employ two numerical methods, exact diagonalization (ED), namely 
full and Lanczos diagonalization, as well as the infinite Time Evolving Block 
Decimation (iTEBD) method \cite{PhysRevLett.98.070201,PhysRevB.78.155117}. 
For ED we use symmetries, total $S_z$ conservation, 
translation by two sites, spin flip for $S_z=0$, and parity combined 
with  translation by one site, to reduce the computational effort. 
In the presence of a uniform magnetic field, the energy levels  
of the system change according to $E_n(h) = E_n(0) \pm hM$ where $E_n(0)$ is  
an eigenvalue of the Hamiltonian in the absence of the magnetic field.  
High  energy states belonging to higher $S_z$ sectors will lower 
their energy in the presence of the magnetic field and will become 
the ground state as the magnetic field reaches a certain value. This process for a finite 
system yields finite steps in the magnetization curve, which are not true 
plateaus but merely finite-size effects. In turn, one needs to discriminate between 
real magnetization plateaus and finite-size effects. 

For a modulated $g$-factor,  the situation becomes numerically more demanding. 
First, the unit cell is enlarged, which creates problems for both methods. Regarding ED,  
the number of $k$-points is reduced, meaning larger Hilbert spaces for each subsector, 
creating a memory threshold at smaller system sizes.  An additional issue 
is that the energy levels in the presence of the magnetic field can no longer be evaluated 
parametrically from the levels without the magnetic field due to the site dependence of 
$g_j$. This means that each value of the magnetic field needs to be evaluated separately, 
leading to a dramatic increase of the computational time for the larger system sizes.  
Regarding iTEBD,  the $g$-factor modulation causes convergence problems 
due to  the larger unit cell. To avoid this, we rely solely on ED for $\delta g>0$.   
  
\subsection*{Sawtooth chain, uniform $g$-factor}\label{sec:uniform}    
To correctly interpret the ED results, we first contrast  the magnetization under a uniform
 magnetic field  of a sawtooth chain for $\fr=0.5,1,2$ obtained via ED for $L=28$ spins 
 to that obtained via iTEBD for an infinite system, Fig.~\ref{fig:magnetization}(a). 
 For all numerical simulations, we assume $h=B$. 
 For the ED results, we plot $M(h)$ as dashed lines, exhibiting  
finite steps. It has been argued that connecting the middle point of these 
magnetization steps reproduces the magnetization curve in the thermodynamic limit. 
In Fig.~\ref{fig:magnetization}(a), we also show the middle points of the magnetization 
steps,  except at $M_s/2$ where the plateau is expected, and we mark its limiting values. 
From the agreement of the points to the iTEBD data, one can safely argue that ED gives 
an excellent qualitative estimate of  $M(L\rightarrow \infty)$. The only exception to that 
is the size of the plateau for $\fr=2$, which  ED tends to overestimate while from the 
iTEBD it seems rather small. To make this visible,  we plot as an inset in panel
 Fig.~\ref{fig:magnetization}(a) the magnetization for $f=2$ only in the highlighted 
region  of the main panel. The disagreement between the two methods 
is attributed to the finite 
size behavior of the gap at elevated magnetic fields, which is rather small at this region 
of the parameter space. 

Let us now describe the distinct features of the magnetization curve 
of the sawtooth chain  for each value of frustration $\fr$. First, for the weakest  
$\jbv = 0.5 \jbbl$ ($\fr=2$), we observe a very steep increase of $M$ at low 
magnetic fields. This reflects the two decoupled-chain limit 
$\jbv\rightarrow 0$ where the upper spins, being loosely coupled with 
the rest of the system, can be very easily polarized. 
One additional point characteristic of the energy scales is 
that $M(h)$ is a concave function of the magnetic field before the plateau 
and a convex function after it. A similar behavior is observed for 
$\fr=1$, with a much wider  plateau also apparent from 
the ED data \cite{doi:10.1142/S0217979208050176}. As the ratio $\jbv = 2 \jbbl$  
is further increased ($\fr=0.5$) the plateau 
still extends to a wide range of magnetic field range but the magnetization 
now displays a convex behavior for $M(h)< M_s/2$, and a concave one for 
$M(h) > M_s/2$. Hence, the sign of the second derivative of the magnetization 
$\mathrm{sgn}[M''(h)]$ provides a very useful criterion for the relative 
strength of the exchange couplings in the system. 
Decreasing further the ratio of the $\jbv/\jbbl$ would lead to a decrease in the size of 
the plateau, since the system comes closer to the non-frustrated Heisenberg chain
\cite{PhysRev.133.A768}.  

In terms of localized magnons for $f=0.5$ (Sec.~\ref{sec:localized}), 
and using Eq.~\eqref{eq:hsat},  we find the saturation field to be 
$\hsat /J= 4 J_2/J=2.4$, for $J=(2 J_1+J_2)/3$,  which is exactly the numerical 
value obtained from both methods.  We also note that at the saturation field, 
there is a complete degeneracy of all localized-magnon  states with energy 
$\varepsilon_n =0$. As a result, $m=M/M_s$ jumps between  the saturation 
value $m=1$ and the value $m=1-\nmax/M_s=1/2$, with $\nmax=N/4$. 
\subsection*{Sawtooth chain, modulated $g$-factor}\label{sec:modulated} 
Now that the ED  has been tested and its results can be correctly interpreted, 
we will use it to study the magnetization process in the presence of a spatially 
varying $g$-factor.  In Figs.~\ref{fig:magnetization}(b) and (c), 
we present results for the magnetization in the presence of a modulated factor 
for different deviations $\delta g$ and for two values of the frustration ratio 
$\fr=0.5,2$, respectively. We observe that in both cases, 
a relatively small deviation of $\delta g/g = 0.2$ already significantly 
extends the plateau region. As $\delta g $ is further increased, the plateau 
grows even more while in the extreme case where $\delta g \geq g$, i.e., a 
$g$-factor with a staggering sign,  the system 
is locked in the half-saturation plateau and never reaches full saturation. 
Classically thinking, this behavior is to be expected because the spins on the sites 
which have a $g$-factor of strength $g$ will polarize faster. However, to satisfy 
the antiferromagnetic interactions of the system, the spins which experience 
a weaker magnetic field will order anti-parallel to the magnetic field to reduce the energy, 
and therefore the $M_s/2$ plateau is favored. Lastly, 
we also observe that the $g$-factor modulation introduced here does not 
affect the sign of $M''(h)$, which seems to depend solely on the exchange 
couplings. 

From Sec.~\ref{sec:localized} and Eq.~\eqref{eq:hsat},  
the prediction for the values depicted in Fig.~\ref{fig:magnetization}(b) is 
that $\hsat/J=2.8$ for $\delta g =0.2$, which is in remarkably good agreement 
with the numerical results. For $\delta g =0.5$, the theoretical prediction 
is $\hsat/J=3.6$, which deviates from the numerical value $\hsat/J= 4.1$, 
suggesting that first-order perturbation theory employed here is insufficient. 
From Fig.~\ref{fig:magnetization}(b) it also becomes apparent that for high values 
of $\delta g$, the degeneracy in $n$ is lifted, and there is a number of critical 
fields $\hcr(n)$ for which the magnetization $M=M_s - n$ changes subsector, 
with $1 \leq n \leq \nmax$. For the value of $f$ chosen in 
Fig.~\ref{fig:magnetization}(c) the picture of localized magnons holds no longer 
and, therefore, no comparison to the theoretical predictions of Sec.~\ref{sec:localized} 
can be made. 

\subsection*{Zig-zag ladder}  \label{sec:zigzag}
Lastly, in the fourth panel of the magnetization data, Fig.~\ref{fig:magnetization}(d), 
we apply the same idea but to the zig-zag ladder  ($\jbbl=\jbbu$), which is known to
exhibit a plateau at $M_s/3$ for a uniform magnetic field \cite{Honecker_2004, 
doi:10.1143/JPSJ.74.1849}.  Although the zig-zag ladder is invariant 
under chain inversion, the $g$-factor modulation considered here breaks chain reflection 
symmetry, enlarging the unit cell of the otherwise translationally-by-one-site invariant 
model to four. As one can see in Fig.~\ref{fig:magnetization}(d), the plateau at 
$M_s/3$ is destroyed in favor of creating a large plateau at $M_s/2$, as predicted 
by the field theory calculation.  

\section{Conclusion}\label{sec:conclusions}

In conclusion, we presented a comprehensive theoretical study of the $J_1-J_2$ 
sawtooth as well as the frustrated chain focusing on their magnetization 
process. A unified field theory for both models was developed, and we demonstrated 
that by introducing a site dependence to the g-factor, the magnetization 
plateau of the sawtooth chain at $M_s/2$ grows for any $J_2/J_1$ ratio
while the $M_s/3$ plateau of the frustrated chain is destroyed in favor of 
a $M_s/2$ plateau. For anisotropies where the g-factor 
vanishes or becomes staggered, we found that the system is locked in the 
$M_s/2$ plateau, never reaching full saturation. We also emphasized  the role 
of the curvature of $M$ for acquiring an estimate of the microscopic couplings. 
We anticipate our results to provide guidelines for future theoretical and 
experimental studies, aiming in new ways to manipulate and extend the plateau 
region in frustrated magnets, including regimes where the 
plateau is expected to be small or even non-existing. 
\section{Acknowledgments}  
We are thankful to Leonie Heinze for useful discussions, Xenophon Zotos for 
his comments on the sine-Gordon model, and Stefan S\"{u}llow for motivating 
this work. Work of W.B. has been supported in part by the DFG through
Project A02 of SFB 1143 (Project-Id 247310070), by Nds.
QUANOMET, and by the National Science Foundation under
Grant No. NSF PHY-1748958. C.P. has received funding from
 the European Union's Horizon 2020 research and innovation 
programme under the Marie Sklodowska-Curie grant agreement No 839004. 
W.B. also acknowledges the kind hospitality of the PSM, Dresden.
%
%
%
%
%
%
%
\end{document}